\documentclass[prb,showpacs,twocolumn,amsmath,amssymb,floatfix,superscriptaddress]{revtex4-1}

\usepackage{graphicx}
\usepackage{pstricks,pst-node,pst-text,pst-3d}
\usepackage{color}

\graphicspath{{Figures_new/}}

\newcommand{\pdagger}{{\phantom{\dagger}}}
\newcommand{\dt}{\Delta\tau}
\newcommand{\Mp}{$\mathbf{M^\prime}$}
\newcommand{\X}{$\mathbf{X}$}
\newcommand{\vMp}{\mathbf{M^\prime}}
\newcommand{\vX}{\mathbf{X}}

\newcommand{\refs}[1]{Sec.~\ref{sec:#1}}
\newcommand{\refss}[1]{subsection \ref{subsec:#1}}

\newcommand{\reff}[1]{Fig.\ \ref{fig:#1}}
\newcommand{\reffa}[1]{Fig.\ \ref{fig:#1}(a)}
\newcommand{\reffb}[1]{Fig.\ \ref{fig:#1}(b)}
\newcommand{\reffc}[1]{Fig.\ \ref{fig:#1}(c)}
\newcommand{\reffl}[1]{Figure\ \ref{fig:#1}}

\newcommand{\myparagraph}[1]{{\it #1} -- }

\newcommand{\vk}{\mathbf{k}}
\newcommand{\neel}{N\'{e}el}

\makeatletter
  \def\l@subsubsection#1#2{}%
\makeatother

\begin{document}
	\title{Momentum-dependent pseudogaps in the half-filled two-dimensional Hubbard model}

\author{D.~Rost}
\affiliation{Institute of Physics, Johannes Gutenberg University, Mainz, Germany}
\author{E.~V.~Gorelik}
\affiliation{Institute of Physics, Johannes Gutenberg University, Mainz, Germany}
\author{F.~Assaad}
\affiliation{Institute of Theoretical Physics and Astrophysics, University of W\"urzburg, W\"urzburg, Germany}
\author{N.~Bl\"umer}
\affiliation{Institute of Physics, Johannes Gutenberg University, Mainz, Germany}

\date{\today}
  \begin{abstract}     
We compute unbiased spectral functions of the two-dimensional Hubbard model by extrapolating Green functions, obtained from determinantal quantum Monte Carlo simulations, to the thermodynamic and continuous time limits. Our results clearly resolve the pseudogap at weak to intermediate coupling, originating from a momentum selective  opening of the 
charge gap.  A characteristic pseudogap temperature $T^*$, determined consistently from the spectra and from the momentum dependence of the imaginary-time Green functions, is found to  match the 
dynamical mean-field critical temperature, below which antiferromagnetic fluctuations become dominant. Our results identify a regime where pseudogap physics 
is within reach of experiments with cold fermions on optical lattices. 
  \end{abstract}
  \pacs{71.10.Fd, 71.27.+a, 71.30.+h, 74.72.-h}
  \maketitle


\section{Introduction}
A peculiar feature of (underdoped) high-$T_c$ superconductors is the 
coupling of  antiferromagnetic fluctuations to charge degrees of freedom, which leads to a strong momentum dependence  of the 
spectral functions. In particular, it induces pseudogaps in the normal state, i.e., a suppression of the density of states at the Fermi energy, which can be probed using (angular resolved) photo\-emission, inverse photoemission, and related techniques.  The pseudogap shares the $d$-wave symmetry with the order parameter of the superconducting phases occurring at low temperatures and near optimal doping.\cite{Marshall1996,Lee2006,Armitage2010}

Pseudogap physics can also be expected in the undoped Hubbard model.  In the 
absence of  electronic correlations, the tight binding  model is characterized by a coherence temperature $T_{\text{coh}}$, set by the bandwidth $W$.
For weak Hubbard interaction $U \lesssim W$,  antiferromagnetic spin fluctuations, with energy scale $T_{\text{spin}}$, will  develop below the coherence  temperature.  Hence, the temperature window  $ T_{N} < T < T_{\text{spin}}  < T_{\text{coh}}$   is characterized by a metallic state coupled to antiferromagnetic fluctuations, which sets the stage for pseudogap physics. 
Here $T_{N}$ is the \neel\ temperature, at or below which long range order generates a full gap in the presence of perfect nesting (in dimensions $d\ge 2$, with $T_N=0$ in $d=2$). 

Theorists have tried to verify this scenario on the basis of numerical simulations and to compute reliable spectra for decades. Direct simulations can only be performed for clusters of finite extent, usually employing periodic boundary conditions. Early determinantal quantum Monte Carlo\cite{Blankenbecler1981} (DQMC) studies at moderately weak coupling ($U/t=4$) led to spectra with significant low-temperature pseudogap features only for small cluster sizes; thus, 
pseudogaps in the undoped $2d$ Hubbard model were regarded as pure finite-size (FS) artifacts.\cite{White1989,White1992} Later studies at similar coupling strengths\cite{Vekic1993,Creffield1995,Moukouri2000,Huscroft2001} 
found pseudogaps also at large cluster sizes, but did not allow for quantitative predictions in the thermodynamic limit. A recent study using the dynamical vertex approximation (D$\Gamma$A) observed reentrant behavior incompatible with the earlier results.\cite{Katanin2009}

A central limitation of all previous numerical pseudogap studies is that results for different cluster sizes (e.g. in DQMC simulations) were compared only at fixed temperatures and at the level of spectral functions. With increasing cluster size, these  show diverse effects: shifts of spectral peaks, transfer of spectral weight, and the opening or closing of gaps. 
A direct {\it pointwise} extrapolation of these positive semidefinite and normalized functions is clearly impossible. In fact, we are not aware of any published attempts of deriving spectral properties in the thermodynamic limit from DQMC data in any context.

In this paper, we present (i) the local spectral function, (ii) momentum-resolved spectral functions at high-symmetry points, and (iii) momentum-resolved spectral functions along high-symmetry lines of the Brillouin zone in the thermodynamic limit. All results are based on FS extrapolations of imaginary-time Green functions, obtained from DQMC, with subsequent analytic continuation to the real axis using the maximum entropy method (MEM)\cite{Jarrell1996} and, in case (iii), a Fourier fit of the momentum dependence. 
The final results are free of significant systematic errors and represent the thermodynamic and continuous time limits in an unbiased way.

Thereby, we can not only unambiguously confirm the pseudogap scenario and study the nodal--antinodal dichotomy in unprecedented detail, but also explore the temperature dependence of the pseudogap opening and identify a characteristic temperature $T^*$.  
At  weak to intermediate  couplings, $T^*$ tracks the onset of  short-ranged magnetic fluctuations, and is equally shown to compare remarkably well with the dynamical mean-field critical temperature for antiferromagnetic long-range order.


In \refs{model}, we introduce the model, set up our notation, characterize the established methods (DQMC, MEM) underlying our approach, and specify our implementations. 
The new methods for eliminating systematic biases from Green function and spectra
are, then, presented in \refs{extrapolation}, first for the DQMC Trotter error, then for finite-size effects.
Our main results are discussed in \refs{results}, starting with pseudogap features in the spectral functions
for the ``nodal'' and ``antinodal'' 
\footnote{The terms ``nodal'' and ``antinodal'' refer originally to the $d$-wave order parameter in high-$T_c$ superconductors, which has a node (vanishing gap) near the point $\mathbf{M'}$ and is maximal near the antinode $\mathbf{X}$ [cf.\ \reffa{BZ} and \reff{A_path}]. The same is true for the pseudogap, which is maximal at $\mathbf{X}$.}
high-symmetry $\vk$ points on the Fermi surface and their evolution as a function of temperature. 
We then show, with continuous momentum resolution, how the pseudogap evolves throughout the Brillouin zone (BZ) and discuss non-Fermi liquid physics that is not accessible by conventional methods. Finally, we determine the characteristic pseudogap temperature $T^*$ for $U\le W$ and relate it to spin correlation functions and other characteristic temperature scales of the model.


\section{Model and conventional methods}\label{sec:model}

\subsection{Hubbard model}\label{subsec:Hubbard}
Our starting point is the single-band Hubbard Hamiltonian with nearest-neighbor hopping $t$ on a square lattice (with unit lattice spacing $a\equiv 1$):
  \begin{eqnarray}\label{Hubb_mod}
    \hat{H} &=& \hat{H}_0 
  \,+\, U \sum_{i} \hat{n}_{i\uparrow}\, \hat{n}_{i\downarrow}\\
\hat{H}_0 \,&=&\, -t\! \sum_{\langle ij\rangle ,\sigma}  \hat{c}^{\dag}_{i\sigma}  \hat{c}^\pdagger_{j\sigma} 
\,=\,  \sum_{\vk, \sigma} \varepsilon_\vk\, \hat{n}_{\vk \sigma}\\[0.5ex]
 \varepsilon_\vk &=& -2 t[cos (k_x) + cos(k_y)] 
  \end{eqnarray}
	\noindent
  Here, $\hat{c}^\pdagger_{i\sigma}$ ($\hat{c}^{\dag}_{i\sigma}$) are annihilation (creation)
operators for a fermion with spin $\sigma\in \{\uparrow,\downarrow\}$ at site $i$; $\hat{n}_{i\sigma} = \hat{c}^{\dag}_{i\sigma} \hat{c}^\pdagger_{i\sigma}$. In the following, the energy scale will be set by $t\equiv 1$.  

At half filling $n\equiv \langle \hat{n}_{i\uparrow} + \hat{n}_{i\downarrow}\rangle =1$ and in the noninteracting limit $U=0$, the occupied momentum states form a square (dark shaded) within the square Brillouin zone illustrated in \reffa{BZ}\ (in the thermodynamic limit), which implies a perfect nesting instability: The Fermi surface (gray line) transforms into itself when shifted by the antiferromagnetic wave vector $(\pi,\pi)$.
As a consequence, any finite interaction $U>0$ drives this model to long-range antiferromagnetic order (only) in the ground state. 

While a conventional notation is well established for the center $\mathbf{\Gamma}$ and the corner $\mathbf{M}$ of the BZ of the square lattice as well as for the antinodal \X\ point, this seems not to be the case for the nodal point; in \reff{BZ} and in the following, we denote this midpoint of $\overline{\mathbf{\Gamma M}}$ as \Mp.

\begin{figure} 
	\unitlength0.1\columnwidth
	\begin{picture}(10,5.1)
	  \put(0.5,-0.3){\includegraphics[height=.543\columnwidth]{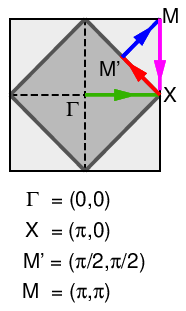}}
      \put(5,0){\includegraphics[height=.5\columnwidth]{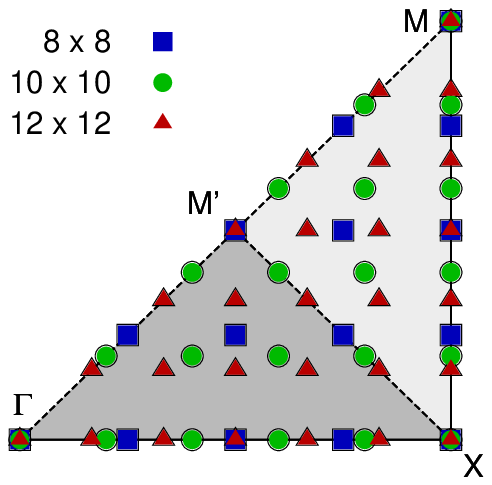}}
	  \put(0,4.7){(a)}
	  \put(4.8,4.7){(b)}
	\end{picture}
	\caption{(Color online) Brillouin zone (BZ) of the square lattice: (a) full BZ with Fermi surface (gray line) and occupied momenta in thermodynamic limit (dark shaded) at $U=0$; letters denote high-symmetry points; color lines and arrows indicate the path used in \reff{A_path}. (b) Irreducible wedge of BZ with momenta occurring in finite-size clusters of linear dimension $L=8,10,12$ with periodic boundary conditions.  \label{fig:BZ}}
\end{figure}

\subsection{Determinantal quantum Monte Carlo algorithm}\label{subsec:DQMC}
The Hubbard model, Eq.\ (1), is solved in this work for clusters with a finite number $N$ of sites [implying a discrete momentum grid, see \reffb{BZ}] at finite temperatures $T$ using the DQMC algorithm developed by Blankenbecler, Scalapino, and Sugar,\cite{Blankenbecler1981} with modifications by Hirsch.\cite{Hirsch1988} It is based on (i) a uniform discretization of the imaginary-time interval $0\le \tau \le \beta$ [with $\beta=1/(k_B T)$], occurring in the path integral, into $\Lambda$ time slices of width $\dt=\beta/\Lambda$, (ii) a Trotter decoupling of kinetic and interaction terms, and (iii) a Hubbard-Stratonovich transformation which replaces the electron-electron interaction at each time slice and site by the coupling of the electrons to a binary auxiliary field. Expectation values are obtained by Monte Carlo importance sampling of field configurations, with weights given by a product of two determinants for the two spin components. In the particle-hole symmetric case considered in this study, this product is always positive, i.e., the sign problem is absent. The numerical effort scales as $\beta N^3$. A detailed review of the algorithm used can be found in Ref.\ \onlinecite{Assaad2008}.

In this work, we obtained imaginary-time Green functions and spin correlation functions between each pair of sites by applying the DQMC method to square lattice clusters $L \times L$ of the linear size $L=8,10,12,14,16$ with periodic boundary conditions, using a set of Trotter discretizations with $0.1\le\Delta\tau\le 0.42$ and typically 50 bins with 5000 sweeps over the auxiliary field each. For the largest systems, individual runs took about a month of computer time; up to five such runs were averaged over in order to reduce error bars. This resulted in typical statistical errors in the (finite-size) Green functions of ${\cal O}(10^{-4})$. Note that the DQMC scaling with $L^6$ makes it difficult to access much larger system sizes directly: $L=20$ ($L=32$) would increase the effort by a factor of about $4$ (64) compared to $L=16$. Local properties were averaged over all sites, momentum ($\vk$) dependent properties were obtained by Fourier transforms of the real-space measurements. 

\subsection{Maximum entropy method}\label{subsec:MEM}
Since DQMC calculations provide Green functions $G$ (and correlation functions) only at imaginary times, their interpretation as dynamical information requires an analytic continuation to the real axis. Specifically, one has to invert relations of the form
\[
G(\tau) = - \int_{-\infty}^\infty d\omega\, A(\omega)\, \frac{e^{-\tau\omega}}{1 + e^{-\beta\omega}},
\]
where $A(\omega)=-\text{Im}\, G(\omega)/\pi$ is the corresponding spectral function. This is an ill-posed problem, as the exponential kernel suppresses the impact of features in $A(\omega)$ at large $|\omega|$ on $G(\tau)$; in the DQMC context, further complications arise from the fact that $G$ is only measured on the discrete imaginary-time grid $\{\tau_l = l \dt\}_{l=0}^{\Lambda-1}$. The MEM finds the most probable spectrum, given the data $\bar{G}_l \pm \Delta G_l$, by balancing the misfit of a given candidate spectrum (measured by the corresponding $\chi^2$) with an entropy constraint which favors smooth spectra.\cite{Gubernatis1991,Jarrell1996} In our implementation, the resulting minimization problem is solved deterministically using a Newton scheme in the singular space of the kernel.
Its application both to DQMC raw data for local and $\vk$ dependent Green functions and to Green functions obtained from Trotter and/or FS extrapolations always resulted in reliable and consistent maximum entropy spectra.


\section{Extraction of unbiased spectra in the thermodynamic limit}
\label{sec:extrapolation}

What sets our main results, to be presented in \refs{results}, apart from earlier work, is their direct relevance in the thermodynamic limit, i.e., the absence of significant bias. We now specify our methodology for eliminating Trotter and finite-size errors from Green functions and establish its accuracy and reliability on the level of Green functions and spectra.

\subsection{Trotter errors and extrapolation $\dt\to 0$}\label{subsec:Trotter}

As discussed in \refss{DQMC}, the DQMC method decouples electronic interactions (and evaluates, e.g., Green functions) at the cost of introducing an artificial imaginary-time discretization $\dt$, which implies an unphysical bias in all DQMC estimates of observables. In the absence of phase transitions, DQMC raw results are expected (and observed) to depend smoothly on $\dt$, in the form of a power series; for some static observables, such as the total energy, it is easy to prove\cite{Fye1986} polynomial dependence on $\dt^2$.

The effects of the Trotter discretization on the 
imaginary-time Green function $G(\tau)$ are illustrated in 
\reffa{dtau_GA}:
\begin{figure}[t] 
	\unitlength0.1\columnwidth
	\begin{picture}(10,4.7)
	  \put(0.1,0){\includegraphics[width=.48\columnwidth]{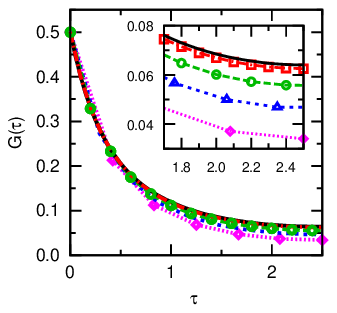}}
    \put(5.2,0){\includegraphics[width=.48\columnwidth]{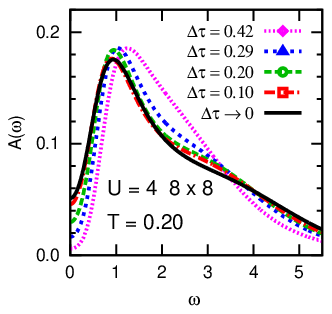}}
	  \put(0,4.1){(a)}
	  \put(5.1,4.1){(b)}
	\end{picture}
	
	\begin{picture}(10,3.9)
	  \put(0.5,0){\includegraphics[width=0.9\columnwidth]{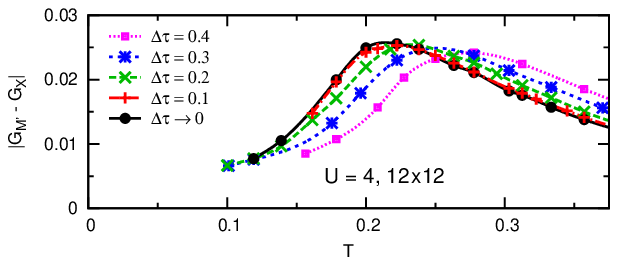}}
	  \put(0.6,3.4){(c)}
	\end{picture}
\caption{\label{fig:dtau_GA} (Color online) 
Impact of Trotter discretization and extrapolation of $\dt\to 0$: 
(a) for local imaginary-time Green functions $G(\tau)$ at $U=4$, $T=0.20$, $L=8$ in the range 
$0\le \tau \le \beta/2=2.5$ (inset: magnified view for $\tau \ge 1.7$); 
(b) corresponding local spectral functions $A(\omega)$.
(c) for the difference between nodal and anti\-nodal Green functions (cf.\ \refss{Ts}) versus temperature.}
\end{figure}
(i) each of the raw data sets (symbols) lives on a different $\tau$ grid; (ii) at fixed values of $\tau$, the data (or a linear interpolation - dashed/dotted lines) is shifted to smaller absolute values at larger $\dt$. Obviously, unbiased results (for a fixed cluster size $L$ in real space) can only be expected after an extrapolation of $\dt\to 0$. On the other hand, such an extrapolation is not possible locally, i.e., at fixed $\tau$, but requires a global approach that can use input from DQMC raw data at all discretizations $\dt$ for each imaginary time $\tau$ of interest.

The black solid lines in \reffa{dtau_GA}\ represent the result of a multigrid procedure, originally developed in the context of the Hirsch-Fye quantum Monte Carlo method for the Anderson impurity model.\cite{Blumer2008,*Gorelik2009} The multigrid method is based on the fact that ``reference'' Green functions $G_{\text{ref}}(\tau)$ with sufficiently accurate asymptotics at $\tau\to 0$ (and $\tau\to \beta$), in particular for the curvature, can easily be derived from weak-coupling expansions (or, alternatively, from the ``best'' QMC data via MEM); consequently, the difference between the measured Green functions $G_{\dt}(\tau_i)$ and the reference $G_{\text{ref}}(\tau_i)$ can be adequately represented by a natural cubic spline for each value of $\dt$; all of these splines can, then, be evaluated on a common (fine) grid.
For this transformed data, we find a nearly linear dependence on $\dt^2$ (plus a small curvature), so that pointwise extrapolations $\dt\to 0$ are reliable and accurate. At the level of $G$, the shift of the unbiased result [black solid line in \reffa{dtau_GA}] of about $10^{-3}$ compared to the best raw data [at $\dt=0.1$, squares and dash-dotted line in \reffa{dtau_GA}] is still significant.

This is no longer true on the level of spectra, shown in \reffb{dtau_GA}, due to the intrinsic complications of MEM: The results for $\dt=0.1$ agree within accuracy with the unbiased spectrum. Thus, we may conclude that $\dt=0.1$ is ``good enough'' for spectral data (at $U=4$) and that an elimination of the Trotter error is not necessary for reducing unphysical bias below significance. At the same time, the smooth consistent evolution of the spectra with $\dt$ confirms our MEM procedure both for the DQMC raw data and for extrapolated Green functions.

Even smaller Trotter errors than observed in \reffa{dtau_GA} can be expected for differences of Green functions, due to error cancellation. Indeed, the scalar pseudogap measure $|G_\mathbf{M'}-G_\mathbf{X}|$, to be introduced in \refss{Ts}, is impacted significantly by Trotter errors only for large discretizations; the bias become negligible for $\dt\lesssim 0.1$, as shown in \reffc{dtau_GA}. Therefore an explicit elimination of this error is, again, not necessary.


\begin{figure} 
\unitlength0.1\columnwidth
	\begin{picture}(10,4.66)
	  \put(0.1,0){\includegraphics[width=.48\columnwidth]{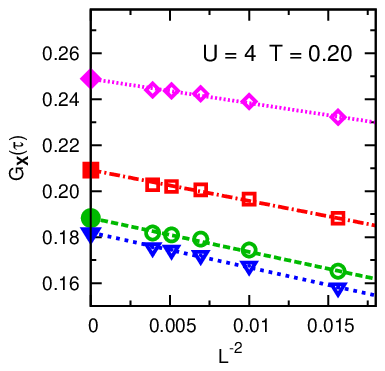}}
    \put(5.2,0){\includegraphics[width=.48\columnwidth]{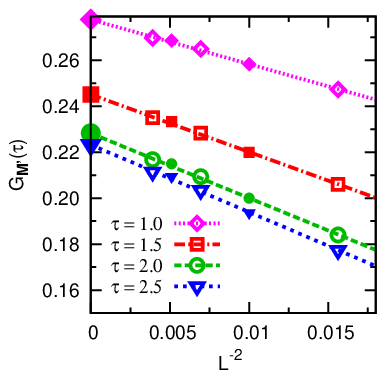}}
	  \put(0,4.3){(a)}
	  \put(5.1,4.3){(b)}
	\end{picture}
	\caption{(Color online) Imaginary-time Green functions (for selected values of $\tau$) versus squared inverse linear dimension (empty symbols) and least-squares extrapolation $L\to\infty$ (lines and large full symbols) at high-symmetry momentum points:  (a) at antinodal \X\ point, (b) at nodal \Mp\ point (small symbols for $L=10,14$: from Fourier fit as shown in \reff{ext_lines}).\label{fig:FS_G_tausel}}
	\label{FSext_GX}
\end{figure}
\subsection{Finite-size extrapolations of local spectra and on high-symmetry $\vk$ points in the BZ} \label{subsec:ext_local}

A FS extrapolation of local properties or $\vk$ resolved properties at high-symmetry points is relatively straightforward: One accumulates raw data at various values of the linear extent $L$ and then extrapolates using polynomial least-square fits in $1/L^2$. In the case of imaginary-time Green functions, independent extrapolations have to be performed for each value of $\tau$ (on the grid with spacing $\dt$ in the case of DQMC raw data or the grid chosen in the extrapolation $\dt\to 0$ discussed in the previous section). As shown in \reff{FS_G_tausel} for $U=4$, $T=0.2$, the Green function depends on system size quite significantly at generic imaginary times [except for the limits $\tau\to 0$ or, equivalently, $\tau\to \beta$ (not shown)] both at the antinodal (a) and nodal (b) momentum points. At the same time, the dependence is quite regular so that least-square extrapolations (lines) can be restricted to low orders. 

Obviously, this ``local'' procedure can only include lattice sizes for which the $\vk$ point under consideration exists [cf.\ \reffb{BZ}]; for the antinodal point, this requirement is fulfilled for all even values of $L$, while the nodal \Mp\ point is only present if $L$ is a multiple of 4 (which restricts the set to  $L=8,12,16$ in our study). Still, as seen in \reffb{FS_G_tausel}, the extrapolation is reliable even with only three data points (per fit), as the dependence is almost perfectly linear. 
\footnote{Additional data points for $L=10,14$ [small symbols in \reffb{FS_G_tausel}] were not included in the FS extrapolations (lines), but are clearly consistent with them; this confirms the accuracy of the Fourier fits of momentum dependencies (see \refss{ext_path}) from which they originate.}
At the same time, the FS extrapolation is particularly important at the nodal \Mp\ point, as FS effects are much stronger than in the antinodal case [shown in \reffa{FS_G_tausel}].

Note that $4\times 4$ clusters (with periodic boundary conditions) have a special symmetry: They have the same topology as a $2\times 2\times 2\times 2$ hypercube with open boundary conditions; as a consequence, the next-nearest neighbors along one of the axes and the ones along the diagonal become equivalent, which implies that the \X\ and \Mp\ points are identical in momentum space at $L=4$. In order to avoid the associated extra bias we exclude this system size and consider only lattices with $L\ge 8$ in this study.

The full resulting Green functions in the thermodynamic limit are shown as solid lines in \reff{FS_GA} for the antinodal (a) and nodal (b) $\vk$ points, respectively, together with their finite-size equivalents (dashed and dotted lines).
\begin{figure} 
\unitlength0.1\columnwidth
	\begin{picture}(10,7.74)
	  \put(0.03,4){\includegraphics[width=0.48\columnwidth]{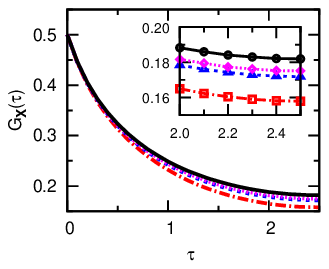}}
    \put(5.2,4){\includegraphics[width=0.48\columnwidth]{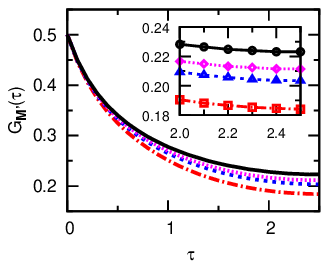}}
 	  \put(0.03,0){\includegraphics[width=0.48\columnwidth]{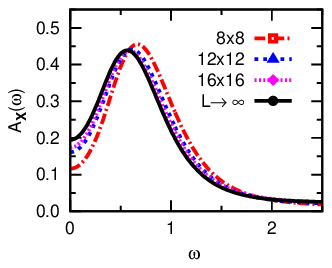}}
    \put(5.2,0){\includegraphics[width=0.48\columnwidth]{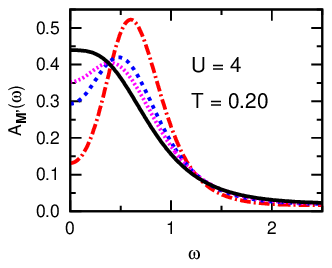}}
	 \put(0,7.5){(a)}
	 \put(5.13,7.5){(b)}
	 \put(0,3.5){(c)}
	 \put(5.13,3.5){(d)}
	\end{picture}
	\caption{(Color online) Upper row: imaginary-time DQMC Green functions ($U=4$, $T=0.2$, $\dt=0.1$) at antinodal (a) and nodal (b) points, respectively, for finite-size clusters (open symbols and broken lines) plus extrapolated (cf.\ \reff{FS_G_tausel}) results in thermodynamic limit (filled circles and solid lines). Lower row: corresponding spectral functions.\label{fig:FS_GA}}
\end{figure}
We see, again, that FS effects are much more prominent at \Mp\ [note the different scales in the insets of \reffa{FS_GA} and \ref{fig:FS_GA}(b)]. The effect is even much stronger on the level of the corresponding spectra, shown in \reffc{FS_GA} and \ref{fig:FS_GA}(d), respectively: In an $8\times 8$ system (dashed-dotted line), nodal and antinodal spectra are qualitatively very similar, with a clear pseudogap feature, and differ mainly in peak height (at $|\omega|\approx 0.7$); at $\vk=\vX$, the spectrum remains nearly unchanged at larger system sizes and in the thermodynamic limit. At $\vk=\vMp$, in contrast, the pseudogap dip shrinks significantly for larger systems and is completely lost in the thermodynamic limit, where a quasiparticle shape appears. This shows that essential pseudogap physics, with a nodal--antinodal dichotomy, is really a property of the thermodynamic limit and that the bias inherent in finite-size systems dangerously distorts the physical picture.


\subsection{FS extrapolations of spectra along high-symmetry momenta in the BZ} \label{subsec:ext_path}

The elimination of FS errors at generic momenta requires ``global'' extrapolations that involve some kind of functional fitting procedures in momentum space. For momenta along high-symmetry lines, these fits have the form of Fourier series which may be adapted in order to take all symmetries into account. In the following, we will illustrate the algorithm for the most important path, the irreducible portion $\overline{\mathbf{X M'}}$ of the noninteracting Fermi surface. This path can be parametrized as
\[
  k_x = (2-\kappa)\, \pi/2;\qquad k_y = \kappa\, \pi/2;\qquad \kappa\in [0,1]\,;
\]
then $\kappa=0$ corresponds to $\mathbf{X}\equiv (\pi,0)$ while $\kappa=1$ corresponds to $\mathbf{M'}\equiv (\pi/2,\pi/2)$. At particle-hole symmetry, all functions $f$ have to be symmetric with respect to both end points, which implies that they can be represented in the form
\[
  f(\kappa) = a_0 + \sum_{n=1}^\infty a_n \sin^2(n\, \kappa\, \pi/2)
\]
with coefficients $a_n$. We have chosen to fit the difference of the Green function for each $\vk$ (along the line) with respect to the antinodal Green function (corresponding to $\kappa=0$); this implies that the zeroth-order coefficient vanishes exactly. The symbols in \reffa{fit_extrapolation_path}\ represent DQMC data for the difference Green functions at $\tau=\beta/2=2.5$;
\begin{figure}[t] 
	\unitlength0.1\columnwidth
	\begin{picture}(10,4.35)
	  \put(0.01,0){\includegraphics[width=.578\columnwidth]{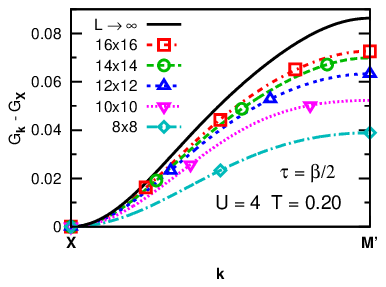}}
    \put(6.02,0){\includegraphics[width=.39\columnwidth]{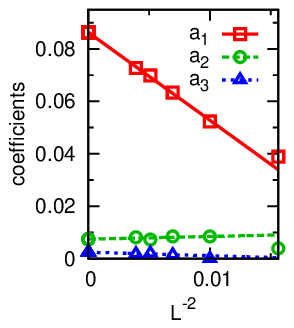}}
	  \put(0,4.0){(a)}
	  \put(6,4.0){(b)}
	\end{picture}
\caption{(Color online) Example of finite-size extrapolation of imaginary-time Green functions along high-symmetry lines in the BZ, here for the path $\mathbf{X} \to \mathbf{M^\prime}$ [cf.\ \reffa{BZ}] and $\tau = \beta /2$: 
(a) difference Green functions with respect to $G_\mathbf{X}$ (symbols), fitted with a Fourier series (broken lines) 
and final result of the extrapolation to the thermodynamic limit (black solid line), 
(b) extrapolation of the corresponding Fourier coefficients. \label{fig:ext_lines}}
\label{fig:fit_extrapolation_path}
\end{figure}
evidently their interpolation using the above Fourier series up to third order (dashed/dotted lines) works quite well. Furthermore, the associated Fourier coefficients depend very regularly (i.e., almost perfectly linearly) on $1/L^2$, as seen in \reffb{fit_extrapolation_path}, and decay exponentially as a function of order. Consequently, an extrapolation to the thermodynamic limit is possible on the level of the coefficients (using a least-squares fit) with high precision; the extrapolated coefficients yield a reliable estimate $G_\vk(\tau=2.5)$ for all $\vk$ along the path [solid line in \reffa{fit_extrapolation_path}]. This procedure has to be performed independently for each value of $\tau$; spectra can then be obtained using MEM on an arbitrarily dense $\vk$ grid.
Similar procedures were employed separately for each high-symmetry line indicated in \reffa{BZ}.

\section{Results}\label{sec:results}

\subsection{Pseudogap signatures at nodal and antinodal $\vk$ points}\label{subsec:results_nodal}

\begin{figure}[t] 
	\unitlength0.1\columnwidth
	\begin{picture}(10,7.7)
	  \put(0,0){\includegraphics[width=\columnwidth]{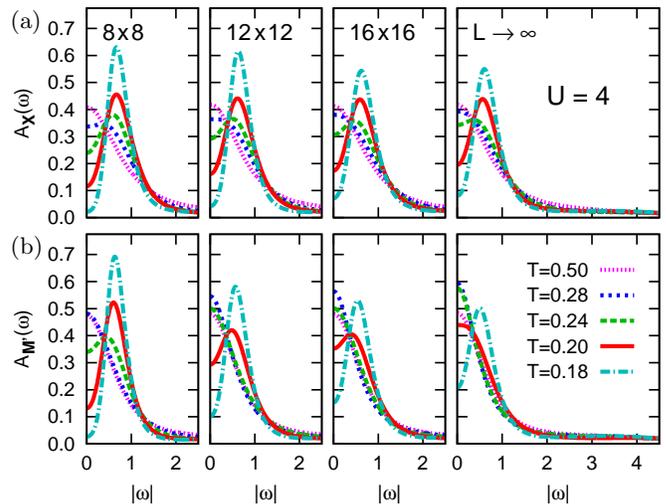}}
	  \put(0,7.3){(a)}
	  \put(0,3.83){(b)}
	\end{picture}
\caption{ \label{fig:spectra_U4}(Color online) Evolution of the DQMC spectral functions with temperature at $U=4$ for finite clusters with $8\le L \le 16$ and in the thermodynamic limit: (a) at the antinodal point [$\vk=\mathbf{X}\equiv (\pi,0)$], (b) at the nodal point [$\vk=\mathbf{M'}\equiv (\pi/2,\pi/2)$].}
\end{figure}

Let us, first, turn to the antinodal and nodal spectra shown in \reffa{spectra_U4} and \ref{fig:spectra_U4}(b), respectively. 
At the elevated temperature $T=0.50$ (dotted lines) the spectra have quasiparticle (QP) shape at all system sizes and for both momentum points. FS effects are negligible: Even the spectra of the smallest systems considered ($8\times 8$, left column) do not deviate visibly from those in the thermodynamic limit (right column); also the momentum dependence along the Fermi surface is minimal at $T=0.50$, with about $20\%$ larger peak height at the nodal \Mp\ point.

In the $8\times 8$ case (left column), the largest system size fully considered in previous studies, a pseudogap dip appears almost simultaneously at $T=0.28$ and $T=0.24$ (dashed lines) at the \X\ and \Mp\ points, respectively, and quickly deepens to an almost complete gap at $T=0.18$ (dashed-dotted line). Given only this data, one would conclude that any momentum dependences beyond the free dispersion are inessential, i.e., that the physics might be in reach of theories with a momentum-independent self-energy [in particular, the dynamical mean-field theory (DMFT)]. However, this picture is distorted by finite-size effects and far from the truth:
In the thermodynamic limit (right column in \reff{spectra_U4}), the antinodal spectra have QP shape only for $T\ge 0.28$; at $T=0.24$, a slight dip emerges at $\omega=0$ which develops to a significant PG at $T=0.20$ and an almost complete gap at $T=0.18$.
\footnote{The evolution of the peak position toward low $T$ is consistent with a ground state charge gap (Ref.\ \onlinecite{Assaad1996}) $\Delta_c\approx 0.67$.}
In contrast, the nodal spectrum retains QP form down to $T=0.20$ (while even the $16\times 16$ system shows a PG dip at this temperature), before a PG emerges at $T=0.18$. Thus, the FS extrapolation detailed above is really essential for fully resolving the nodal--antinodal dichotomy, which is at the heart of PG physics at finite temperatures. Only in the limit $T\to 0$, i.e., in the presence of long-range order, one expects a DMFT-like picture to become valid (again, as for high $T$) with fully gapped spectra all along the Fermi surface.

This implies that finite-size effects should mainly have two consequences on the level of spectra: (i) shift characteristic PG temperatures upwards, (ii) dilute the nodal-antinodal dichotomy in the vicinity of these characteristic temperatures.

\myparagraph{Characteristic PG temperature $T^*$} As the opening of the PG is not a thermodynamic phase transition, 
it lacks a unique critical temperature. It is still useful (and common)\cite{Marshall1996,Lee2006,Armitage2010,Vidhyadhiraja2009} to define a {\it characteristic PG temperature} $T^*$, for comparison with other temperature or energy scales of the system. 
An obvious choice of the required scalar PG measure is a dip in the spectral function. We specify this ``pseudogap strength'' by the reduction of spectral weight at $\omega=0$, compared to the maximum: 
\[
 r_\vk \equiv 1 - A_\vk(\omega=0) / \max_{\omega} A_\vk(\omega)\,,
\]
as shown for the (anti)nodal points in \reff{PG_strength}.
\begin{figure}[t] 
	\unitlength0.1\columnwidth
	\begin{picture}(10,3.8)
	  \put(0,0){\includegraphics[width=\columnwidth]{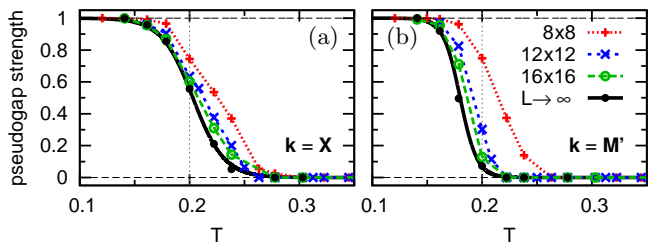}}
	  \put(4.7,3.1){(a)}
	  \put(5.9,3.1){(b)}
	\end{picture}
\caption{ \label{fig:PG_strength}(Color online) Scalar measure of pseudogap strength (see text) versus temperature: (a) at antinodal, (b) at nodal point. The nodal--antinodal dichotomy is fully apparent only in the thermodynamic limit (solid lines).}
\end{figure}
This representation reveals that the onset of the PG is slow only at $\vk=\mathbf{X}$: As soon as $r_\mathbf{X}\approx 0.5$, $r_{\mathbf{M'}}$ jumps to the full value within a narrow temperature range $\Delta T\approx 0.03$. The results in the thermodynamic limit (filled circles) can be fitted with a Fermi function form (solid lines); using their inflection points yields $T^*_\mathbf{X} \approx 0.20$, $T^*_{\mathbf{M'}}\approx 0.18$. Note that, again, the FS effects are much stronger at $\vk=\mathbf{M'}$ than at $\vk=\mathbf{X}$.

\myparagraph{Comparison with the literature}
In a pioneering study, Huscroft {\it et al.}\cite{Huscroft2001} had obtained first bounds on the FS errors in DQMC spectra
by complementing DQMC results for $N\le 64$ sites with those of
the dynamical cluster approximation (DCA) employing $N$ $\vk$ patches in the self-energy. 
The resulting anti\-nodal
spectral functions for $U=5.2$ are shown as dashed and dotted lines in \reffa{spectra_U52}, respectively.
\begin{figure}[t] 
	\unitlength0.1\columnwidth
	\begin{picture}(10,4.7)
	  \put(0,0){\includegraphics[width=\columnwidth]{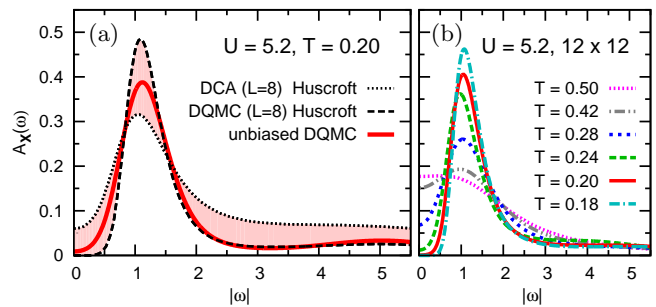}}
	  \put(1.2,4.1){(a)}
	  \put(6.47,4.1){(b)}
	\end{picture}
  \caption{ \label{fig:spectra_U52} (Color online) Spectral functions of the half-filled Hubbard model at the antinodal point [$\vk=\mathbf{X}\equiv (\pi,0)$] and for weak coupling $U=5.2$:
 (a) Unbiased spectrum at $T=0.20$ (solid line), in comparison with earlier DCA and finite-size DQMC results.\cite{Huscroft2001} (b) DQMC spectra for $12\times 12$ lattice.
}
\end{figure}
The shaded region denotes the bounds in which one would expect the true spectrum, according to the opposite FS tendencies (with DQMC over- and DCA underestimating gaps at small cluster sizes) of both methods. Note that the remaining uncertainty is still significant and that the bounds are not rigorous, due to numerical noise and the difficulties of the MEM.

Our unbiased estimate of $A_\mathbf{X}(\omega)$, shown as solid line in \reffa{spectra_U52}, reduces these uncertainties drastically: We find that the spectral weight at low frequencies ($|\omega|\lesssim 0.3$) is  much smaller than predicted by DCA, but still significant (i.e., larger than the raw DQMC prediction). The true peak height at $|\omega|\approx 1.1$ is close to the average of the DCA and DQMC predictions. At large frequencies $|\omega|\gtrsim 1.5$, we find excellent agreement with the earlier DQMC estimates\cite{Huscroft2001} which shows that the DQMC FS errors are small in this region (cf.\ \reff{spectra_U4}) and also verifies the procedures for analytic continuation; in contrast, DCA is still far off (at $N=64$).

Compared with the results for $U=4$ presented in \reff{spectra_U4}, our unbiased result [solid line in \reffa{spectra_U52}] shows much stronger PG characteristics, as is certainly expected at the stronger interaction $U=5.2$. Spectra for a full range of temperatures at this interaction are shown in \reffb{spectra_U52} for a $12\times 12$ system; these results can directly be compared with the second column in \reff{spectra_U4}.
Already at the highest temperature $T=0.50$ [dotted line in \reffb{spectra_U52}], the spectral peak is much broader, i.e., more spectral weight has been shifted away from the origin than at $U=4$. This tendency towards more insulating behavior remains at lower $T$: The peak-to-peak width is about twice as large as for $U=4$. At $T=0.18$ (dash-dotted line), no spectral weight can be resolved at $|\omega|\lesssim 0.5$, so that the PG looks numerically like a full gap. In addition, the characteristic PG temperature is clearly shifted upwards, with a well-developed PG already at $T=0.28$; the dependence of $T^*$ on $U$ will be studied more broadly in \refss{Ts}. 

\subsection{Evolution of pseudogap in full momentum-resolved spectral function}\label{subsec:results_full}

So far, we have presented results which, for given parameters $U$ and $T$, are of a similar nature as those previously discussed in the literature. The main advances of our study of nodal and antinodal spectra are (i) our elimination of the (enormous) finite-size bias inherent in raw results and (ii) our explicit analysis of temperature effects. We will now turn to fundamentally new results, namely spectra with full momentum resolution.

\begin{figure} 
	\includegraphics[width=\columnwidth]{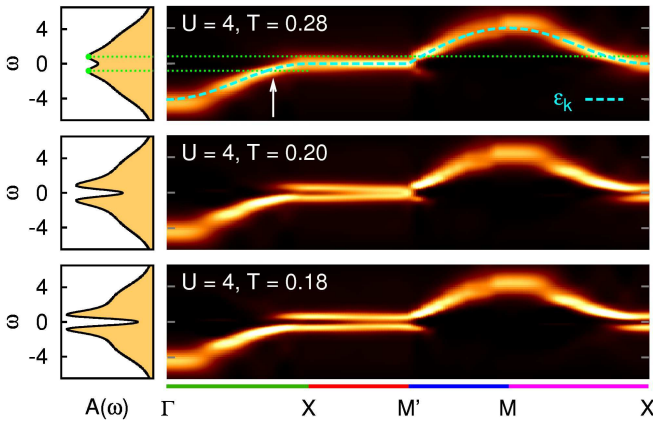}
	\caption{(Color online) Unbiased local spectra $A(\omega)$ (first column) and momentum-resolved spectra $A_\vk(\omega)$ for $U=4$ and $\vk$ along the path through the Brillouin zone illustrated in \reffa{BZ}; the pseudogap opens with strong $\vk$ dependence at $T\le T^*\approx 0.20$. A local maximum in the spectral density at $\omega\not=0$ (arrow) is indicative of spin-polaron physics. 
	\label{fig:A_path}}
\end{figure}

\reffl{A_path} shows unbiased momentum-resolved spectra $A_\vk(\omega)$ throughout the whole Brillouin zone, along the path indicated in \reffa{BZ}, at weak coupling $U=4$ and in a temperature range $0.18\le T\le 0.28$; in addition, the left column contains the local spectra $A(\omega)$, corresponding to an average over all $\vk$. 
We have chosen a path $\overline{\mathbf{\Gamma X M' M X}}$ that contains the irreducible portion $\overline{\mathbf{X M'}}$ of the noninteracting Fermi surface (at half filling).
The inclusion of this subpath allows us to study the nodal--antinodal dichotomy continuously and in detail; more generally, all variations along this path (where $\varepsilon_\vk=0$) arise from a $\vk$ dependent self-energy, i.e., effects beyond DMFT. 

At $T=0.28$ (first row in \reff{A_path}), the local intensity maxima are unique at each $\vk$ point
and agree rather well with the noninteracting dispersion $\varepsilon_\vk$ (dashed line), except for the edges $\omega\gtrsim 4$.
A well-defined quasiparticle peak at $\omega\approx 0$ (especially sharp near $\vk=\mathbf{M'}$ and more diffuse at $\vk=\mathbf{X}$) is consistent with a Fermi liquid description. This picture changes at $T=0.20$ (second row), when the spectrum splits at $\vk\approx \mathbf{X}$, i.e., a pseudogap opens at the antinodal point, while the rest of the spectrum (at momenta with $\varepsilon_\vk \not=0$) is essentially unchanged. The gap size decreases smoothly on the line $\mathbf{X}\to \mathbf{M'}$. Only at $T\le 0.18$ (third row) the QP is destroyed also at $\vk=\mathbf{M'}$; a PG then extends over all momenta. 

Compared to the strong temperature dependence along the path $\overline{\mathbf{X M'}}$, the spectra appear nearly unchanged in the rest of the BZ. In particular, a sharp dispersive quasiparticle like band, indicated by an arrow in the top panel, evolves from the \X\ point about half way towards the $\mathbf{\Gamma}$ point (and, equivalently by particle-hole symmetry, from the \X\ point towards the ${\mathbf{M}}$ point). We interpret this feature, which is not accessible in conventional DQMC studies at FS, as the formation of a spin polaron band (arrow in \reff{A_path}), with an energy offset at lower $T$ indicating the magnetic exchange scale. It ends (at higher $|\epsilon_\vk|$) in a ``waterfall'' which breaks up the band structure into low and high energy features.\cite{Preuss1997}

Taken together, our results indicate that, apart from incoherent features at $|\omega|\approx 4$ which are present at all temperatures and should continuously evolve into Hubbard bands with increasing $U$, interaction effects come into play with lowering $T$ first very locally (in momentum space) around the antinodal \X\ point; apparently, the strong enhancement of scattering by the van Hove singularity at \X\ completely determines the physics in this region. This explains why the spectra can become sharper, implying a reduction in the imaginary part of the self energy, on the path from \X\ towards $\mathbf{\Gamma}$ (up to the position of the arrow in \reff{A_path}, corresponding to the energy $\omega$ indicated by dotted lines), i.e., with increasing $\varepsilon_\vk$ and $\omega$; a behavior which is exactly opposite to usual Landau Fermi liquid and also to DMFT physics.

This suppression of spectral weight around \X\ already at elevated temperatures also explains the slight dip seen in the local spectrum at $T=0.28$ (dots and dotted lines in top panel of \reff{A_path}; cf.\ also \reff{MEM}): 
\begin{figure}[tb] 
\includegraphics[width=\columnwidth]{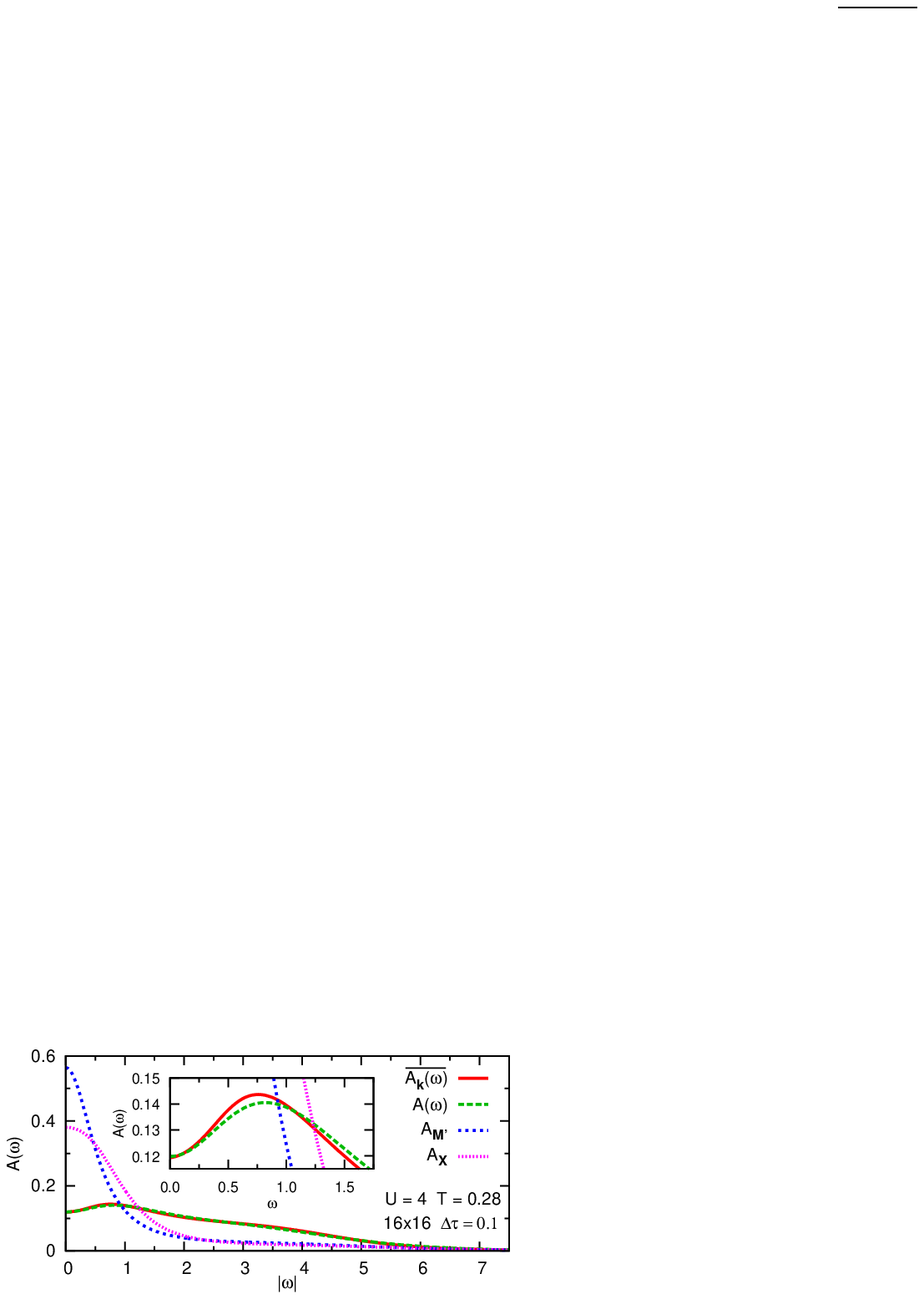}
\caption{ \label{fig:MEM}(Color online) Test of MEM accuracy at $U=4$, $T=0.28$: 
The local spectral function $A(\omega)$ (dashed line), calculated from the local Green function 
$G(\tau)$, agrees well with the average  $\overline{A_\vk(\omega)} \equiv \frac{1}{N} \sum_\vk A_\vk(\omega)$ (solid line). Both curves reveal a dip at $\omega\lesssim 0.8$ which is absent at this temperature in $A_{\mathbf{M^\prime}}(\omega)$ 
(short-dashed line) and $A_{\mathbf{X}}(\omega)$ (dotted line). 
}
\end{figure}
While the momenta around \Mp\ and in the spin-polaron band region (arrow) contribute ``normally'' to the local spectrum, the contributions from momenta near \X\ are spread out to about a much larger width (with a significant fraction at $|\omega|\gtrsim 1$); the missing weight at $\omega\ll 1$ results in the dip. 

One might worry that this analysis puts too much confidence in the accuracy of our data and that the dip in the local spectrum at $T=0.28$, a local suppression in $A(\omega)$ by abound $10\%$ in a narrow frequency range, corresponding to a ``missing weight'' of about $1\%$, could also result from uncertainties in the MEM procedure. Therefore, we have checked its consistency and accuracy in the largest finite-size system ($16\times 16$) by comparing the local spectrum $A(\omega)$ (dashed line in \reff{MEM}), obtained by direct analytic continuation using MEM from the local Green function $G(\tau)$ with the average of all (here 256) momentum-resolved spectra $A_\vk(\omega)$ in \reff{MEM}. As $G(\tau) \equiv \frac{1}{N}\sum_\vk G_\vk(\tau) \equiv \overline{G_\vk(\tau)}$, both spectra should agree, if evaluated exactly: $A(\omega) \stackrel{!}{=} \overline{A_\vk(\omega)} \equiv \frac{1}{N}\sum_\vk A_\vk(\omega)$. As the MEM is inherently nonlinear, due to the entropy constraint, deviations must be expected in practice. However, our procedure, with very accurate DQMC data, seems to be quite stable: 
Although the $\vk$ dependent spectral functions $A_\vk(\omega)$ differ substantially at different $\vk$ points (shown in \reff{MEM} only for the nodal and antinodal points using short-dashed and dotted lines, respectively) and have much more pronounced features than the local spectral function $A(\omega)$ (long-dashed line), their average $\overline{A_\vk(\omega)}$ (solid line) agrees with it nearly within linewidth; only the magnified inset reveals tiny differences at small frequencies. So we conclude that our techniques are more than adequate and that the small dip discussed above is, indeed, physical.

Let us, finally, stress that our eliminations of finite-size 
errors have been absolutely essential for obtaining unbiased momentum-resolved 
spectra, as illustrated in \reff{density_FE} for the path $\mathbf{X} \to \mathbf{M'}$: 
Not only is the convergence at the end points $\vk=\mathbf{X}$ and $\vk=\mathbf{M'}$ slow, the  
$\vk$ resolution is also quite coarse, with only one intermediate point for $L=8$ 
and only three intermediate points for $L=16$. It is clear that a very significant 
extension of the cluster size (e.g. to $64\times 64$, implying a factor of 
$4^6=4096$ in computer time) would be needed in order to match the momentum 
resolution of our extrapolation procedure. 
\begin{figure}[t] 
	\includegraphics[width=\columnwidth]{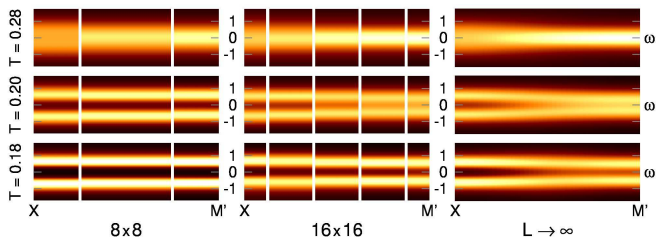}
	\caption{ \label{fig:density_FE}(Color online) Spectral functions $A_\vk(\omega)$ 
for $\vk$ along the Fermi edge (line from $\mathbf{X} \to \mathbf{M^\prime}$; cf.\ \reff{BZ}). FS results ($L=8,16$) converge (slowly) to the thermodynamic 
limit both point-wise and by refinements of the $\vk$ resolution.}
\end{figure}

\subsection{Evolution of characteristic pseudogap temperature $T^*$ with interaction $U$}\label{subsec:Ts}

Apart from yielding a momentum dependent $T^*$, the criterion used in \refss{results_nodal} has the disadvantage of depending on the ill-conditioned analytic continuation of the imaginary-time DQMC Green functions to the real axis. On the other hand, it is difficult to define specific PG criteria on the level of the imaginary-time  Green functions [cf.\ \reffa{FS_GA} and \ref{fig:FS_GA}(b)].
\footnote{The observable $\beta\, G(\beta/2)$ gives also hints about PG physics (Refs.\ \onlinecite{Werner2009,Gull2009,Gull2010}), but is not a sharp PG criterion: Its value changes only by some $20\%$ between the curves, e.g. for $12\times 12$ and $L\to\infty$ in the inset of \reff{spectra_U4}, although the former corresponds to PG and the latter to QP behavior
.} 
However, the nodal--antinodal dichotomy, i.e., the {\it momentum dependence} of the Green functions along the line $\mathbf{X}\to \mathbf{M'}$ (arising from a momentum dependence of the irreducible self-energy) turns out to be illuminating:
\reffa{spin}\ shows that the norm of the difference between the imaginary-time Green functions,
\[
	|G_{\mathbf{M'}}-G_\mathbf{X}| \equiv \Big\{\int_0^\beta d\tau \big| G_{\mathbf{M'}}(\tau) - G_\mathbf{X}(\tau) \big|^2/\beta\Big\}^{1/2}\,,
\]
is strongly enhanced (at $U=4$) in the temperature range where the PG opens. Not surprisingly, this peak becomes sharper and shifts towards lower $T$ in the thermodynamic limit; the position of the maximum yields a natural unique definition of the characteristic PG temperature $T^*\approx 0.20$, indicated by a vertical dotted line in \reff{spin}.

\begin{figure}[t!] 
\unitlength0.1\columnwidth
\begin{picture}(10,4.1)
  \put(0,0){\includegraphics[width=\columnwidth]{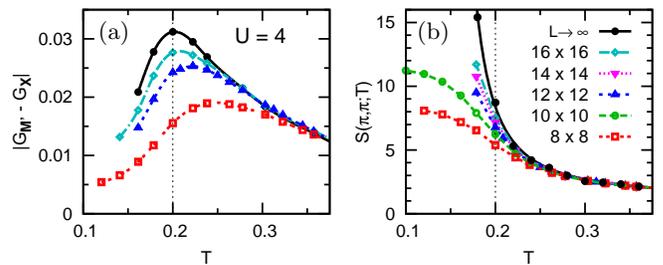}}
  \put(1.4,3.5){(a)}
  \put(6.3,3.5){(b)}
\end{picture}
\caption{ \label{fig:spin}(Color online) Properties of finite clusters and in the thermodynamic limit at $U=4$: (a) Difference between nodal and anti\-nodal Green functions versus temperature; its maximum defines the characteristic PG temperature $T^*\approx 0.20$. (b) Unnormalized spin structure factor at antiferromagnetic wave vector $\vk=(\pi,\pi)$. Dotted vertical lines mark $T^*$.}
\end{figure}

As discussed in the introduction, the PG is associated (at $n\approx 1$) with AF correlations and may be regarded as a precursor of a fully gapped long-range ordered AF phase which, in $d=2$, is realized only in the ground state.\cite{Assaad1996} Thus, we should expect to see a strong enhancement in suitable spin correlation functions. While the nearest-neighbor spin correlations are only very moderately enhanced at $T\lesssim T^*$ (not shown), the spin structure function is seen in \reffb{spin}\ to increase by a full factor of 4 in the range $0.9\,T^*\le T\le 1.1\,T^*$. At the same time, FS effects explode at $T\lesssim T^*$. All this shows that the PG is driven by the development of AF order at a scale which is large compared to the lattice spacing. 

The PG physics and, in particular, the momentum dependence observed at $U=4$ should disappear at strong coupling, when already the high-temperature phase is gapped at $n=1$.
\footnote{The onset of strong AF correlations should still be visible in the higher-frequency portions of the spectral function as it leaves signatures in the kinetic energy (Ref.\ \onlinecite{Gorelik2012}) and optical conductivity (Ref.\ \onlinecite{Taranto2012}).}
The dichotomy should also vanish in the limit $U\to 0$, 
where the energy scale $T_{\text{spin}}$ vanishes, and so does the magnitude of the pseudogap.
Indeed, the momentum dependence is seen in \reff{Gk} 
to peak at $U\approx 4$ and to decay quickly for larger couplings, where also FS effects (which can be estimated from the thin lines, corresponding to $8\times 8$, in comparison to the main $12\times 12$ results) become irrelevant. At fixed cluster size, also the results at weaker coupling ($U=3$, $U=2$) fall off; unfortunately, they suffer from significant FS effects which are too costly to eliminate.
Still, the peak positions allow us to estimate $T^*(U)$ in the full range of weak to intermediate coupling as denoted by symbols in the inset of \reff{Gk}.
\footnote{Our estimate of $T^*\approx 0.30$ at $U=6t$ agrees well with a recent DCA result (Ref.\ \onlinecite{Vidhyadhiraja2009}).}
\begin{figure}[t!] 
\includegraphics[width=\columnwidth]{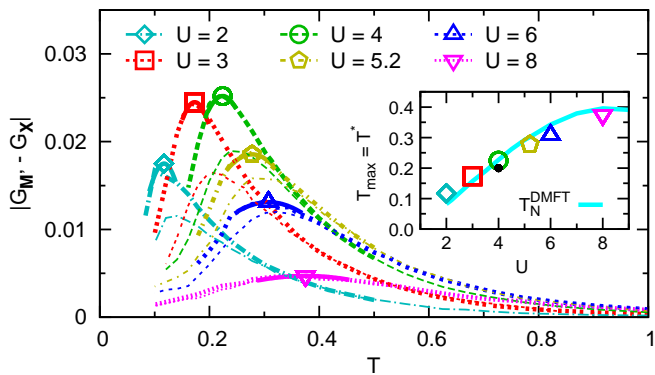}
\caption{ \label{fig:Gk}(Color online) Difference between nodal and anti\-nodal Green functions versus temperature: DQMC results at weak to intermediate coupling ($2\le U \le 8$) for $12\times 12$ clusters (thick lines) and $8\times 8$ clusters (thin lines). Inset: associated $T^*$ in comparison with DMFT {\neel } temperature.}
\end{figure}
Also shown is the mean-field estimate of the critical temperature for AF long-range order (solid line). At first sight, this DMFT estimate of the \neel\ temperature $T_{\text{N}}^\text{DMFT}$ would appear irrelevant, as the true $T_{\text{N}}=0$ by the Mermin-Wagner theorem. However, we find that $T^*\approx 0.9\, T_{\text{N}}^\text{DMFT}$ for $4\le U \le 8$; a correction of FS effects for $U=2$ and $U=3$ should push the corresponding values of $T^*$ also below $T_{\text{N}}^\text{DMFT}$. So the DMFT identifies the relevant temperature scale for spin coherence (as was previously observed in the strong-coupling regime\cite{Gorelik2012}); however, it lacks the momentum resolution which is essential to capture the pseudogap physics explored in this paper.

\section{Conclusion}
After decades of research, our understanding of the two-dimensional Hubbard model, especially regarding the extent to which it captures the pseudogap and high-$T_c$ physics of cuprates, is still far from complete.
Numerical simulations\cite{Werner2009,Gull2009,Vidhyadhiraja2009,Lin2010,Gull2010,Sordi2011,Gull2012} give valuable hints, but continue to be dominated by finite-size effects.
\footnote{In particular, the recent prediction of a Fermi liquid -- superconductor crossover at weak coupling (Ref.\ \onlinecite{Gull2012}), based on paramagnetic DCA (Ref.\ \onlinecite{Maier2005}) with only 8 (or 16) $\vk$ patches, appears inconsistent with our unbiased results.}
We have overcome the finite-size barrier and presented momentum-resolved spectral functions in the thermodynamic limit, obtained by systematic extrapolation of DQMC Green functions ($L\to \infty$ and $\dt\to 0$). Based on this  achievement, we were able to disentangle the delicate interplay of dynamical and spatial magnetic correlations.   
At weak to intermediate couplings,  this interplay leads, indeed, to the formation of a pseudogap in the half-filled band.  The pseudogap  originates from a strong $\vk$ dependence of the self-energy, which results in a $d$-wave-like  anisotropy in the opening of the charge gap and a ``waterfall'' substructure of the spectrum. The associated temperature scale $T^*$ is determined by the onset of antiferromagnetic fluctuations (and nearly agrees with $T_{\text{N}}^\text{DMFT}$), i.e., is rather high compared to other coherence scales and should be in reach of experiments with ultracold fermions on optical lattices.\cite{Hofstetter2002}$^,$ 
\footnote{For pseudogaps in ultracold Fermi gases (without optical lattices) near unitarity, see, e.g., Refs.\ \onlinecite{Gaebler2010,Magierski2011,Tsuchiya2011,Perali2011}.}

\section*{Acknowledgments}

We thank G.\ Sangiovanni for valuable discussions. Financial support by the Deutsche Forschungsgemeinschaft through FOR 1346 and, in part, through SFB/TR 49 is gratefully acknowledged.

\bibliography{DQMC_pseudogap_cite_ab}
\end{document}